# Why does ice float?

Chang Q Sun

Nanyang Technological University, Singapore

ecqsun@Ntu.edu.sg; ecqsun@qq.com

**Abstract**

**The O:H-O segmental specific heat ratio $\eta_L/\eta_H$ defines not only the phase of Vapor, Liquid, Ice I and XI phase with a quasisolid phase that shows the negative thermal extensibility but uniquely the slope of density of water ice in different phases. Ice floats because H-O contracts less than O:H expands in the QS phase at cooling.**

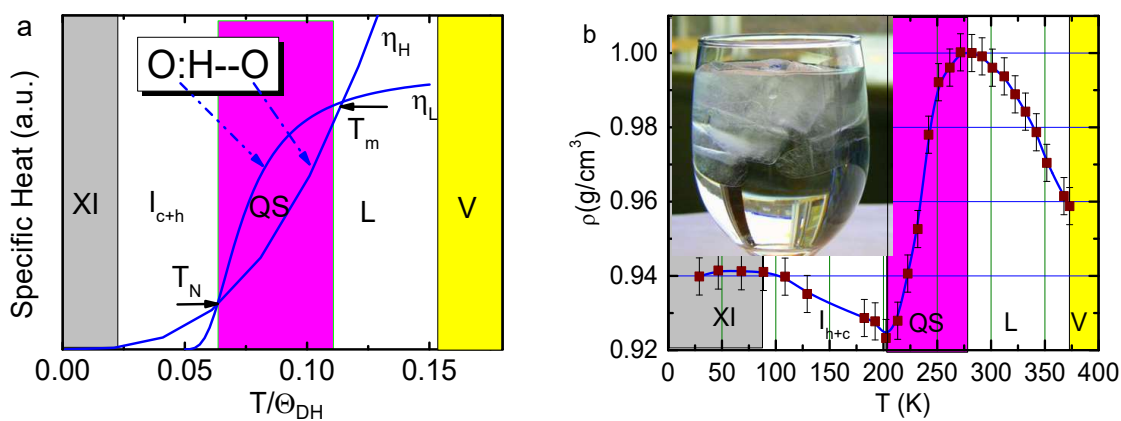

Key Refs:

1. Sun C Q, et al, Density and phonon-stiffness anomalies of water and ice in the full temperature range [J]. *J. Phys. Chem. Lett.*, 2013, 4: 3238-3244.
2. Huang Y L, et al. Hydrogen-bond relaxation dynamics: Resolving mysteries of water ice [J]. *Coord. Chem. Rev.*, 2015, 285: 109-165.
3. Sun C Q and Sun Y, *The Attribute of Water: Single Notion, Multiple Myths*. 2016: Springer-Verlag. 525.

# 1 Introduction

It has been a long-lasting query why ice is less dense than water that makes ice float. A fruitlessly debating between Galileo and Columbus was conducted in 1611 in Florence, Italy, from perspectives of mass density, object shape, surface tension and other possibilities [1]. In 2013, more than 20 scholars from around the world gathered Florence for one week to mark the 400[th] anniversary of the debating [2]. It is still uncertain why does ice float though dozens of hypothetic mechanisms have been proposed. One of the most populations is that water contains domain-resolved low- and high-density phases. It was suggested that the four-coordinated water forms the high-density phase and the chain like structure forms the low-density phase [3]. Computation using TIP4P/2005 water [4] and TIP4P/Ice [5] package suggested recently that mechanical compression raises the density of the four-coordinated $H_2O:4H_2O$ unit cell by squeezing one more molecule into this cell to form the second denser $H_2O:5H_2O$ phase [6]. Cooling raises the fraction of the low density phase, resulting in the occurrence of ice floating, [7,8] in both situations [3,6].

When dealing with the abnormal physical properties of water and ice started from 2010 [9], we formally activated the concepts of the hydrogen bond (O:H-O) as an asymmetrical, short-range, coupled oscillator pair and its segmental specific heat disparity. Coupling the traditionally independent segment has integrated the intermolecular O:H nonbond and the intramolecular H-O polar-covalent bond through the repulsion between electron pairs ":" of the adjacent oxygen atoms, which features properly the cooperative relaxability of the two segments under physical perturbation. At the same time, we implemented the O:H-O segmental specific heat as a function of the cohesive energy and vibration frequency. Thus, the O:H-O segmental disparity and cooperativity in length, energy, vibration frequency and the derived specific heat and polarization have enabled a resolution to multiple anomalies shown by water and ice when subjecting to perturbation by mechanical, thermal, electrical fields and molecular undercoordination [9].

## 2 The coupled hydrogen bond (O:H-O)

Fig.1 inset shows the configuration and the segmental cooperative relaxation under physical perturbation of the coupled O:H-O bond with H being the coordination origin [14]. At the atmospheric pressure and 4 °C temperature, its segment length, energy, and vibration frequency are (0.1 nm, 0. 2 eV, 200 cm$^{-1}$)$_L$ and (0.17 nm, 4. 0 eV, 3200 cm$^{-1}$)$_H$. Subscripts L and H represent O:H and H-O segments, respectively. The coupled hydrogen bonds are subject to the following regulations:

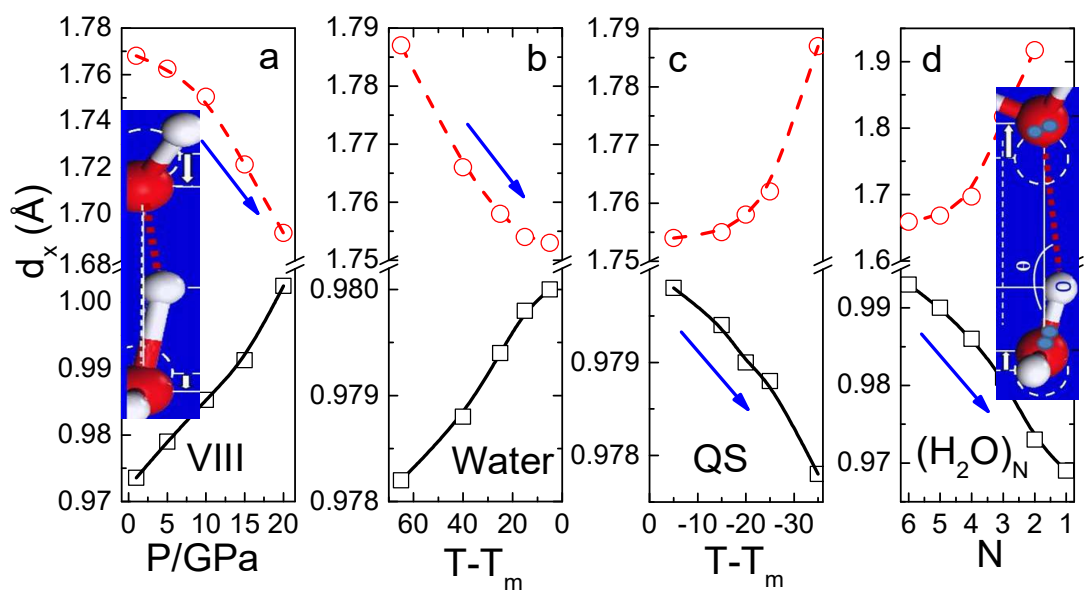

Fig.1. COMPAS force field [28] derived O:H–O cooperative relaxation under stimulus of (a) mechanical compression [29], (b) Liquid cooling, (c) QS cooling [27], and (d) $(H_2O)_N$ cluster molecular undercoordination [30]. Arrows denote the master segments and point to their relaxation directions. **The O:H always relaxes more than the H–O in opposite slopes and curvatures, irrespective of the stimulus applied or the structural phase because of the persistence of O—O repulsive coupling** [14]. Electrification has the same effect of molecular undercoordination, resulting in the supersolidity [31].

1. **Conservation of $sp^3$ orbital hybridization and O:H-O configuration**. Due to the hybridization of $sp^3$ electron orbits of an oxygen, each $H_2O$ molecule has two protons and two pairs of lone pairs, which form the bond between adjacent oxygen, $O^{2\delta-}:H^{\delta}-O^{2\delta-}$ with $\delta \approx 0.6$ - 0.7 e varying with coordination conditions. For convenience in discussion, the δ is taken as a unit. Within a wide temperature and pressure range, the numbers of protons and lone pairs and the configuration and orientation of the coupled hydrogen bonds conserve. Oxygen retains its $sp^3$ electron orbits in the $10^{-11}$ - $10^{12}$ Pa pressures and $10^0$ - $10^3$ K temperatures [15-17]. Even at very high pressures and temperatures, the $2H_2O \leftrightarrow H_3O^+:HO^+$ - super-ionization transition only reverse the orientation of one of the every four O:H-O bonds without de-hybridizing the $sp^3$ orbits [16,17].

2. **Restriction for molecular rotation and proton tunneling**. The rotation of water molecules and the random tunneling of protons at the asymmetric position between neighboring oxygens [18,19] are subject to energy restrictions. Water molecules occupy the center and four apical sites of a $H_2O:4H_2O$ tetrahedron to form four oriented O:H-O bonds. Rotating along the $C_{3v}$ axis of the central $H_2O$ by more than 60 ° will derive the H↔H and O:⇔:O repulsion, which destabilizes the structure. It was found that rotating a $H_2O$ molecule by 120 ° in two-dimensional ice causes the overall structural long-range disorder (known as Bjerrum defect) [20,21]. In addition, the proton translational tunneling only occurs upon H-O broken, while breaking the H-O bond of $H_2O$ in vapor phase (5.1 eV) needs to be excited by a laser radiation of 121.6 cm$^{-1}$ wavelength [22].

3. **Rules for energy exchange and dissipation.** Hydrogen-bonded networks exchange energy with the environment through H-O bond relaxation. H-O bond contraction absorbs and stores energy, while its elongation releases energy. Liquid heating H-O bond contraction has been the foundation of the Mpemba observation of warm water freezing more quickly [23-25]. The network dissipates energy through O:H thermal fluctuation or even molecular evaporation, which consumes energy limited to its 0.2 eV cohesion. Compared with the H-O energy of 4.0-5.1 eV, thermal energy dissipation of one O:H nonbond is only 5% of the energy of one H-O bond. The H-O bond energy rather than the O:H dictates the high heat capacity of the hydrogen bonding network of water and ice.

4. **Hydrogen bond polarization and segmental length cooperative relaxation**. The O:H-O bond can be approximated as an asymmetric, short-range, strongly coupled oscillator pair. Due to the O—O repulsive coupling, a physical perturbation dislocates the oxygen at both ends of the O:H-O bond in the same directions along their connecting line. The displacement of oxygen at the left end of O:H is always greater than the oxygen at the right end of the H-O, $|\Delta d_L| > |\Delta d_H|$, which is accompanied by lone pair polarization except for heating that depolarizes by fluctuation. O:H-O segmental length cooperative relaxation obeys the following relations, where ρ is the mass density [14]:

$$\begin{cases} d_{OO} = 2.6950\rho^{-1/3} & (Molecular\ separation) \\ \dfrac{d_L}{d_{L0}} = \dfrac{2}{1+exp[(d_H - d_{H0})/0.2428]}; & (d_{H0} = 1.0004\ and\ d_{L0} = 1.6946\ at\ 4\ °C) \end{cases}$$

(1)

5. **Hydrogen bond potential and segment vibration frequency**. The three-body potential includes the asymmetrical, short-range O:H Van der Waals potential, H-O exchange interaction, and O—O repulsive coupling. The London dispersion (dipolar interaction with its polarization field), nuclear quantum effect, spin-spin and spin-orbit coupling, and long-range actions are taken as a uniform background. The harmonic approximation of the O:H-O potential function and the segmented vibration frequency solved using the Lagrangian-Laplacian transformation are expressed as:

$$\begin{cases} 2V \cong k_L u_L^2 + k_H u_H^2 + k_C (u_L - u_H)^2 & (\text{Potential energy}) \\ \omega_x \propto \sqrt{(k_x + k_C)/\mu_x} \propto \sqrt{E_x/\mu_x}/d_x & (\text{Phonon frequemcy}) \end{cases} \quad (2)$$

Where $k_x$ and $k_C$ are force constants that are curvatures of the respective potential function for the x-segment and Coulomb coupling at their equilibrium. The $u_x$ is the vibration displacement and $\mu_x$ the reduced mass of the x oscillator. One can directly get the segmental length and vibration frequency by computation and spectroscopy and then transform them into the respective force constant and cohesive energy. Thus, one can obtain the evolution path of the coupled hydrogen bond under physical perturbation. Involvement of the nonlinear contribution to the harmonic approximation of the potential function does not add any spectral feature of bond formation but only shifts spectral peaks slightly [26]. An action of perturbation relaxes the system constantly form its initial to a new equilibrium modulating the bond length and energy.

6. **Specific heat, mass density, and phase diagram.** According to Einstein's relation, the Debye temperature $\Theta_{Dx}$ of a segment is proportional to its characteristic vibration frequency $\omega_x$. On the other hand, the integral of the specific heat $\eta_x(t)$ corresponds to the cohesive energy $E_x$. Thus, the measured segmental vibration frequency and cohesive energy determine uniquely the specific heat:

$$\begin{cases} \Theta_{Dx} \propto \omega_x & (\textit{Einstein relation}) \\ E_x = \int_0^{T_{vx}} \eta_x(t) dt & (\textit{Cohesive enenrgy}) \end{cases} \quad (3)$$

The upper limit of integration $T_{Vx}$ is the critical temperature for the x segment thermal rupture. $T_{VL}$ is the evaporation temperature of the O:H segment at standard atmospheric pressure. Most strikingly, a physical perturbation that specific heat curve through the $\omega_x$ and $E_x$ relaxation. From the known $\Theta_{DL}$ = 192 K of water ice and the O:H stretching vibration frequency 200 cm$^{-1}$, one can derive the $\Theta_{DH} \approx 3100$ K based on the characteristic frequency of the H-O bond at 3200 cm$^{-1}$. Fig. 2a shows that the O:H-O segment specific heat curves have two intersections, which correspond to the mass transition from its density maximum at 277 K closing to the melting point ($T_m$) at 273 K to its density minimum at 258 K for homogeneous ice nucleation $T_N$ [27]. Compared with Figure 2b, it is found that the characteristic points of the specific heat curves correspond to boundaries of the known Vapor, Liquid, $I_h$, $I_c$ and XI phases under atmospheric pressure, in addition to the quasisolid (QS) showing cooling expansion. Because of the high fraction of undercoordinated molecules, the 1.4 nm sized droplet shifts the $T_N$ from 258 to 205 K [8]. Indeed, perturbation by molecular undercoordination modulates the critical temperatures for phase transition through O:H-O relaxation that offsets the specific heat curves.

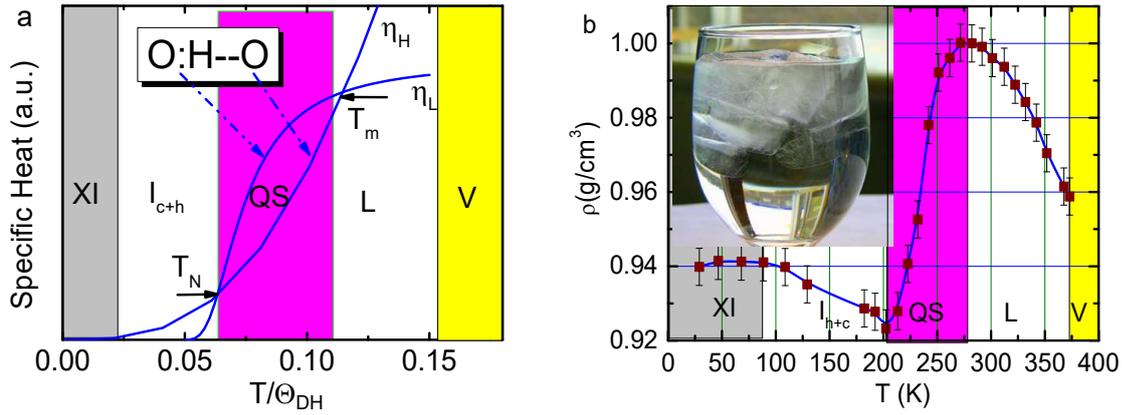

Fig. 2 O:H—O segmental specific heat derived structure phase and mass density variation [27]. Subjecting to cooling expansion, the QS phase boundaries correspond to the extreme densities and close the ($T_N$, $T_m$) for bulk at (258, 277) K. The segmental having a lower specific heat follows the regular rule of thermal expansion while the other segment does it contrastingly. The QS boundary is retractable by external field. The 1.4 nm droplet has the least density and freezes at 205 K [8].

## 2 O:H-O bond cooperative relaxability

### 2.1 Segmental length

Fig.1 shows the cooperativity of O:H-O bond segmental length under perturbation of mechanical compression (P), Liquid (T > 0 °C) and Quasisolid (QS) (T < 0 °C) cooling, and molecular undercoordination of $(H_2O)_{N\leq 6}$ clusters [14] calculated using the COMPASS *ab initio* force field [28]. The segmental length relaxation of the coupled O:H-O occurs in the "master-slave" manner. The arrows are close to the master segments and point to the directions of the applied fields. No matter whatever fields, the O—O distance $d_{oo}$ variation is always realized by one segment extension of the hydrogen bond and the other segment contraction complying with eq (1). For instances, compression or liquid cooling shortens the O:H "master" and the H-O serves as the "slave" undergoing elongation by the O—O repulsive coupling. In contrast, Quasisolid cooling or molecular undercoordination shortens the master H-O while the slave O:H expands cooperatively and contrastingly. Electrification has the same effect of molecular undercoordination on the O:H-O segmental length cooperativity. These observations evidence the essentiality of the O—O repulsive coupling of the O:H-O bond that relaxes in a conventionally unexpected manner.

### 2.2 Vibration frequency

Raman scattering and infrared transmission spectroscopy are the most powerful tools of probing the bond stiffness $Yd$ that is proportional to the square of vibration frequency, $Yd \propto E/d^2 \propto \omega^2$ with $Y \propto E/d^3$ and $d$ being the elastic modulus and bond length, respectively. The spectroscopy gathers the oscillators of the same vibration frequency or identical force constant into a peak, regardless of their spatial positions or orientations in real space based on Fourier transformation. The perturbation of the applied field only changes the abundance of the phonon (spectral peak integral), stiffness (frequency), and structural order fluctuation (full width at half maximum, FWHM). Unless under extreme conditions, a physical perturbation does not create new chemical bonds within the specimen of examination. The relaxation of the phononic characteristics caused by the change of the intensity of the applied field can be filtered by the differential phonon spectrometrics (DPS). The DPS is the difference between of the spectral peaks collected under the action of different field strengths upon the areas of all spectral peaks being normalized. The purpose of spectral peak area normalization is to eliminate artifacts of experimental and system errors. Integral of the DPS peaks above the horizontal axis corresponds to the abundance transition from the reference benchmark (field strength is zero) to the state of being relaxed.

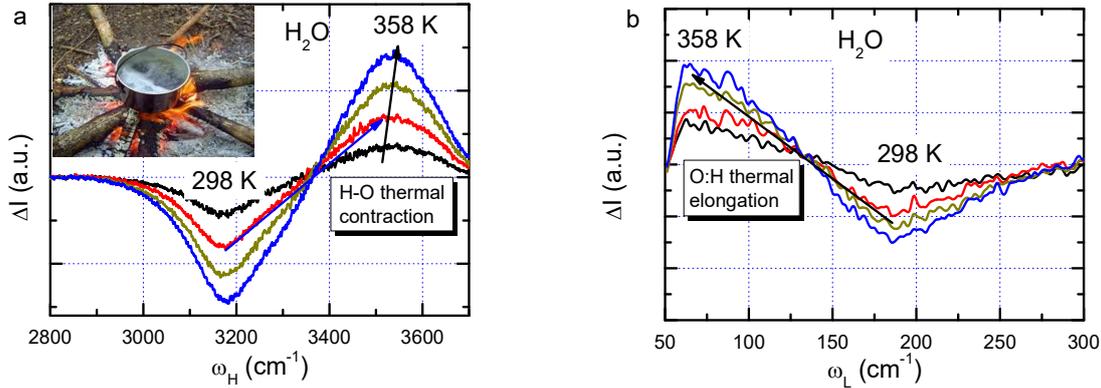

Fig. 3  DPS profiles for water under (a, b) heating [32].

Fig. 3 compares the O:H-O bond phonon relaxation under perturbation of heating, compression, and the skins of water and ice [35] of undercoordinated molecules. Heating stiffens the H-O phonon from 3200 to 3500 cm$^{-1}$ but softens the O:H phonon from 200 to 75 cm$^{-1}$, showing the strong O:H-O segmental cooperativity. The H-O bond thermal contraction absorbs and stores energy and cooling expansion does it reversely, which laid the foundation for the occurrence of Aristotle-Mpemba effect – warmer water freezes more quickly [23-25]. The vibration frequency of the H-O and O:H segments shifts always at the opposite direction regardless of the nature of perturbation.

## 3  O:H-O thermodynamics: why does ice float?

### 3.1  Specific heat ratio versus mass density

Inspecting the specific heat and mass density shown in Fig 2, one can find that the interplay of the segmental specific heat curves not only matches the phase boundaries in the phase diagram but also defines a QS phase bounded at -15 and 4 °C, as confirmed using phonon spectroscopy [27]. The shift of the $T_N$ from -15 to -68 (205 K) °C for the 1.4 nm sized droplet [8] results from the effect of molecular undercoordination. Furthermore, the ratio of specific heat $\eta_L/\eta_H$ defines the rate of density variation in each phase,

$$\frac{d\rho}{dt} \propto \frac{\eta_L}{\eta_H} - 1 = \begin{cases} < 0, & (Liquid) \quad [4, 100\ °C] \\ > 0, & (QS) \quad [-15, 4\ °C] \\ < 0, & (Ice\ I) \quad [-173, -15\ °C] \\ \cong 0 & (Ice\ XI) \quad [<-173\ °C] \end{cases} \qquad (4)$$

At any point of temperature, there are two specific heat values, the segment having a lower specific heat obeys the regular rule of thermal expansion and cooling contraction, while the other segment relaxes reversely because of O—O repulsive coupling. *Fig 4* a shows the segmental length variation with the change of temperature, which determines the mass density, $\rho \propto (d_H + d_L)^{-3}$. The

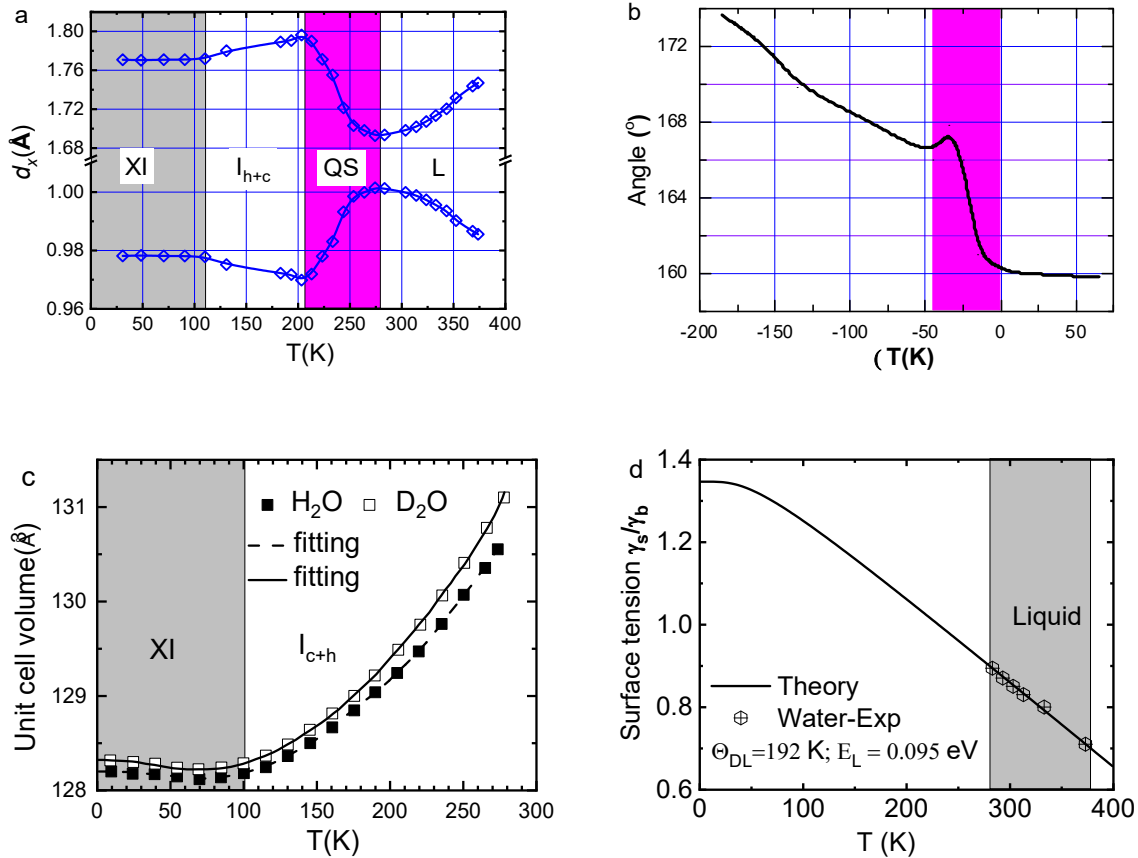

Fig 4  Temperature dependence of (a) O:H-O bond segmental length and (b) ∠O:H-O containing angle variation for bulk water and 1.4 nm sized droplet (T ≤ 273 K reference), (c) volume change of $H_2O$ and $D_2O$ in phase XI (T ≤ 100 K) and phase $I_{c+h}$ [38,39], and (d) thermal depression of liquid surface stress [40].

In the liquid and solid phase I, density decreases when temperature rises, $\eta_L/\eta_H < 1$ and $d\rho/dt < 0$; heating elongates O:H more than H-O contraction, resting on heating volume expansion or cooling contraction. The volume thermal expansion and cooling contraction of Liquid the ice I seemed to be normal, but its mechanism is quite irregular. In the low-temperature XI phase, the segmental length and energy are insensitive to temperature because of the near-zero value of Debye specific heat, so the phonon frequency remains almost constant [41]. As a result of cooling ∠O:H-O angle expansion from 167 to 173° in the XI phase, the density varies slightly, see *Fig 4*b and c [27].

## 3.2 Quasisolid phase cooling expansion

Now let us turn to the QS phase to be updated to the authoritative phase diagram. In the QS phase, the $\eta_L/\eta_H$ ratio reverses from that for liquid and ice I. The H-O of lower $\eta_H$ undergoes cooling contraction and heating expansion while the O:H responds to thermal excitation contrastingly. Because the H-O bond relaxes less than the O:H does, cooling volume expansion or heating volume contraction takes place, which results in the density loss at cooling and gain at heating. At the same time, cooling enlarges the ∠O:H-O angle from 160 to 167°, which adds another factor of the QS phase volume cooling expansion. Therefore, floating ice only occurs intrinsically in the QS state. The QS density drops from its maximum of 1.0 g/cm$^3$ at 4 °C to its minimum of 0.92 at -15 °C. In phase I, the density picks up to 0.94 g/cm$^3$ at its I-XI boundary under cooling. Therefore, the segmental specific heat disparity not only defines the density anomalies in various phases at the ambient pressure but also dictates the cooling volume expansion of the QS phase.

On the other hand, thermal excitation depolarizes electrons and depresses the surface stress. Simulation of the surface stress depression in terms of Debye thermal decay of the skin energy density, shown in *Fig 4* d, turned out the Debye temperature of 192 K and the O:H cohesive energy of 0.095 eV [40] compared with its bulk value of 0.22 eV [42].

## 4 Conclusion

We readily show that the segmental specific heat driven O:H-O bond thermal relaxation govern the mass density of water ice and the ratio of the $\eta_L/\eta_H$ defines uniquely the slope of density of water ice in different phases. Ice floats because of the $\eta_L/\eta_H < 1$. H-O contracts less than O:H expands in the QS phase at cooling so ice floats.